%% file: main.tex
\documentclass[a4paper,11pt]{article}
\usepackage{jinstpub} 
\usepackage{lineno}
\usepackage{amsmath,amsthm}
\usepackage{graphicx,tabularx}
\usepackage{color}
\usepackage{multirow}
\usepackage{comment}
\usepackage{enumitem}
\usepackage{subfigure}
\usepackage{orcidlink}
\usepackage{cleveref}
\usepackage{slashed}
\usepackage[normalem]{ulem}
\usepackage{scalerel}
\usepackage{wrapfig}
\usepackage{cancel}
\usepackage{xcolor}
\usepackage{multirow}
\usepackage{setspace}
\usepackage{dcolumn}	
\usepackage{bm}	     	
\usepackage{setspace}
\usepackage{siunitx}
\usepackage{datetime}
\usepackage{fancyvrb}
\usepackage{textgreek}
\usepackage{xspace}
\usepackage{appendix}
\usepackage{adjustbox}
\usepackage{tikz}
\usetikzlibrary{shapes.geometric, arrows}
\usepackage[normalem]{ulem}

\crefname{section}{Sec.}{Secs.}
\crefname{figure}{Fig.}{Figs.}
\crefname{equation}{Eq.}{Eqs.}
\crefname{table}{Table}{Tables}
\crefname{appendix}{Appendix}{Appendices}

\dedicated{\copyright~2025 CERN for the benefit of the FASER Collaboration. Reproduction of this article or parts of it is allowed as specified in the CC-BY-4.0 license. }

\title{Reconstruction and Performance Evaluation of FASER's Emulsion Detector at the LHC}

\collaboration{\includegraphics[height=17mm]{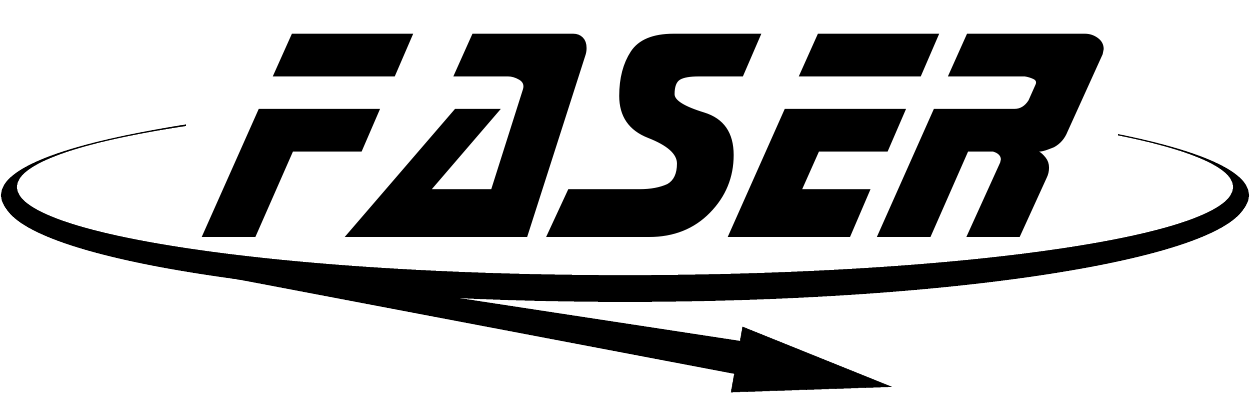}\\[6pt]
FASER Collaboration}

\input{authorlist}

\emailAdd{faser-publications@cern.ch}

\abstract{This paper presents the reconstruction and performance evaluation of the FASER$\nu$ emulsion detector, which aims to measure interactions from neutrinos produced in the forward direction of proton–proton collisions at the CERN Large Hadron Collider. 
The detector, composed of tungsten plates interleaved with emulsion films, records charged particles with sub-micron precision. 
A key challenge arises from the extremely high track density environment, reaching $\mathcal{O}(10^5)$ tracks per cm$^2$. 
To address this, dedicated alignment techniques and track reconstruction algorithms have been developed, building on techniques from previous experiments and introducing further optimizations. 
The performance of the detector is studied by evaluating the single-film efficiency, position and angular resolution, and the impact parameter distribution of reconstructed vertices. 
The results demonstrate that an alignment precision of $\SI{0.3}{\micro m}$ and robust track and vertex reconstruction are achieved, enabling accurate neutrino measurements in the TeV energy range.}


\begin{document}
\maketitle
\flushbottom

\section{Introduction
\label{sec:introduction}}

One of the primary physics goals of the ForwArd Search ExpeRiment (FASER)~\cite{Feng:2017uoz, FASER:2018ceo, FASER:2018bac, FASER:2022hcn} at CERN's Large Hadron  Collider (LHC)~\cite{evans2008lhc} is to study neutrinos produced in the forward direction of the LHC proton-proton ($pp$) collisions, with  energies in the TeV range~\cite{FASER:2019dxq,  FASER:2020gpr}. 
In 2021, the FASER Collaboration reported the first evidence of neutrino interaction candidates produced at the LHC~\cite{FASER:2021mtu}. 
In 2023, the first observation of collider muon neutrinos was achieved using FASER’s electronic detector components~\cite{FASER:2023zcr}. 
Subsequently, the FASER Collaboration reported the first measurement of the muon neutrino interaction cross section and flux as a function of energy~\cite{abraham2024first}. 
The interactions of all three neutrino flavours can be measured by identifying leptons generated through neutrino charged-current (CC) interactions with the FASER$\nu$ emulsion detector~\cite{FASER:2022hcn}. 
For the first time, FASER measured $\nu_e$ and $\nu_\mu$ interaction cross sections using a $\SI{128}{kg}$ subset of the FASER$\nu$ detector after exposure to 9.5~fb$^{-1}$ of 13.6~TeV $pp$ collisions~\cite{FASER:2024hoe}. 
Charged particle tracks are reconstructed with sub-micron precision, enabling the measurement of the energy of the electrons and the momentum of the charged particles, including muons. 

The FASER$\nu$ detector is placed 480 meters downstream from the ATLAS interaction point (IP1), aligned with the collision axis line of sight (LOS). 
Many of the charged particles from the LHC collisions are deflected by the LHC magnets, while neutral hadrons are absorbed by the approximately 100 meters of rock in front of the detector. 
However, given the 40 MHz collision rate at the LHC, many charged particles, such as muons, still reach the FASER$\nu$ emulsion detector.
It is therefore unavoidable to accumulate a large number of charged particle tracks, which poses a challenge for track and vertex reconstruction from neutrino interactions.


The identification of track segments from fully automated emulsion film scanning systems was initially performed by means of the NETSCAN software~\cite{Kodama:2002dk}, originally developed for the DONuT experiment~\cite{DONuT:2007bsg} and later adapted for the CHORUS experiment~\cite{CHORUS:2007wlo}. The FEDRA emulsion reconstruction framework~\cite{Tyukov:2006ny} and NETSCAN 2.0~\cite{Hamada:2012zaa} were subsequently developed for the OPERA experiment. These tools incorporated film-to-film alignment, track reconstruction, topological decay search, momentum measurement, and electromagnetic shower reconstruction. 
A comprehensive historical overview of nuclear emulsion technologies and their development is presented in Ref.~\cite{ariga2020nuclear}.

Recent experiments, such as NA65/DsTau~\cite{DsTau:2019wjb} and FASER, present new challenges, requiring the reconstruction of data in a high track density environment ($\mathcal{O}(10^5)-\mathcal{O}(10^6)$ tracks per cm$^2$), approximately three orders of magnitude higher than OPERA’s track density of 100–1000 tracks per cm$^2$. 
Dedicated reconstruction algorithms to address these challenges were initially developed for DsTau~\cite{DsTau:2019wjb}. 
The reconstruction of the FASER data builds on the FEDRA framework and DsTau algorithms, incorporating further optimizations to suit the specific conditions of FASER.
Alignment techniques were also developed to fully exploit the high spatial resolution of the emulsion detector.


This paper describes the reconstruction techniques developed for FASER$\nu$ and the performance evaluation of the emulsion detector. 
\cref{sec:detector} presents the detector, the temperature monitoring during exposure, and the automated scanning system. 
\cref{sec:reconstruction} describes the reconstruction process, which includes alignment techniques, track reconstruction, and vertex reconstruction. 
\cref{sec:detector_performance} discusses the performance evaluation of the emulsion detector based on the reconstructed tracks.

\section{The Emulsion Detector \label{sec:detector}}

\begin{figure}
    \centering
    \includegraphics[width=0.7\textwidth]{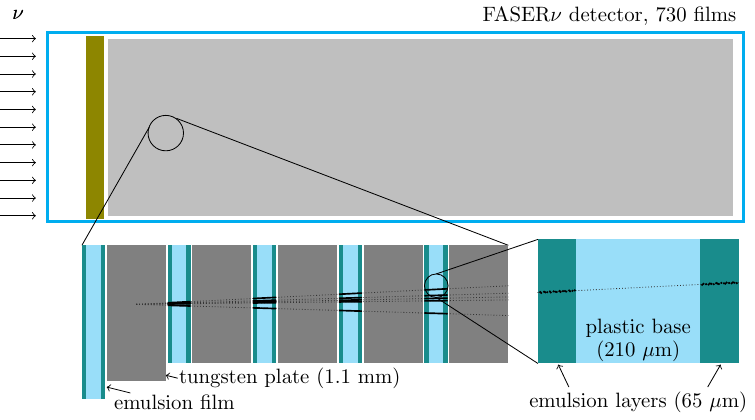}
    \caption{A schematic illustration of the FASER$\nu$ emulsion detector. }
    \label{fig:fasernu_detector}
\end{figure}

\cref{fig:fasernu_detector} shows a schematic of the FASER$\nu$ emulsion detector\footnote{The coordinate system associated to the FASER$\nu$ emulsion detector has the $z$-axis pointing along the LOS away from IP1, the $y$-axis pointing vertically upward, and the $x$-axis pointing horizontally toward the LHC machine. The origin of this coordinate system is aligned with the center of the emulsion detector in the transverse ($x$–$y$) plane and is located at the front of the emulsion detector along the $z$-axis.}.
The emulsion detector has a transverse size of $\SI{25}{\centi\meter} \times \SI{30}{\centi\meter}$ and a length of approximately 100 cm.
The total mass of the detector is approximately \SI{1.1}{\tonne}, corresponding to 220 radiation lengths along the beam. 
It consists of a repeated structure of 730 layers, each composed of a tungsten plate with a thickness of \SI{1.1}{\milli\metre} and an emulsion film. 
The film is made by coating both sides of a transparent polystyrene base with a gel consisting of silver bromide crystals embedded in gelatin~\cite{ariga2020nuclear}. 
The film consists of two emulsion layers, each \SI{65}{\um} thick, deposited on both sides of the \SI{210}{\micro\metre} thick base.

The emulsion gel and the films were produced at the large-scale facility of Nagoya University~\cite{rokujo2024nuclear}. 
In order to enhance sensitivity and reduce background, the films were exposed to high-humidity conditions, 95\% R.H. (Relative Humidity), at $\SI{30}{\degreeCelsius}$ for 18 hours, and then dried for 3 hours at 50\% R.H. at $\SI{25}{\degreeCelsius}$ before detector assembly \cite{Ozaki:2015xlw}. 

The detector assembly was performed at the CERN darkroom facility, with controlled environmental conditions (R.H. of 50\% and temperature of \SI{20}{\degreeCelsius}). 
The detector is organized in 73 modules, each consisting of 10 emulsion films and 10 tungsten plates, which were vacuum-packed with aluminum-laminated foils to isolate the detector from the external environment. 
Atmospheric pressure keeps the relative position of the different films within each module fixed during exposure.

All 73 modules are installed in a mechanical structure, which presses all modules to one side for optimal alignment. 
Since the emulsion detector records all charged particles, it must be periodically replaced.
In fact,  according to the expected muon background flux at FASER and taking into account that the maximum acceptable track density is $10^6$ per $\si{cm^2}$, the detector is replaced two or three times  a year.
Details of the detector and of the operational procedures are reported in Ref.~\cite{FASER:2022hcn}. 

After installation, the emulsion detector is kept at a stable temperature of $\SI{16.5} {\degreeCelsius}$ using a dedicated cooling system. 
The low temperature minimizes the increase in the number of random background silver grains per unit of emulsion area  that develop without particle interaction (fog density). 
Furthermore, the temperature variation is also kept within $\pm \SI{0.05} {\degreeCelsius}$ to avoid thermal expansion of the detector, which could affect the alignment of the emulsions in the FASER$\nu$ box.

After the exposure, the FASER$\nu$ detector is extracted, and the emulsion films undergo specific chemical development in the darkroom facility at CERN. 
This process amplifies the latent image stored in the silver bromide crystals by growing metallic silver filaments, forming visible dots that can be observed under optical microscopes. The amplification gain reaches approximately $10^{8}$, depending on the precise control of the temperature and development time~\cite{FASER:2022hcn}.




The FASER$\nu$ analysis is based on the readout of the entire set of emulsion films by the Hyper Track Selector (HTS) system at Nagoya University~\cite{Yoshimoto:2017ufm}. 
During the scanning process, the films are placed on an acrylic plate and then scanned by the HTS, which is equipped with 72 2-mega-pixel high-speed cameras. It takes 22 tomographic images, and 16 successive images in the emulsion layer are used for track recognition. 
The system has a total data throughput of approximately 50 GBytes/s with 10 GBytes/s effectively used for the image readout. 
The data are distributed to 36 PCs with a total of 72 GPUs. The tomographic image data are processed, and track segments in the emulsion layers, called micro-tracks, are reconstructed in real-time. 
Only micro-tracks are saved, while the raw data images are discarded. The angular acceptance for tracks is set to $\tan\theta<1.5$, where $\theta$ is defined as the track angle to the normal vector to the film surface. 
The reconstruction of micro-tracks is limited by the computing power. While increasing the angular acceptance reduces the scanning speed, the system, in principle, has the potential to detect tracks with $\tan\theta\sim4.0$.

\begin{figure}
    \centering
    \includegraphics[width=0.35\linewidth]{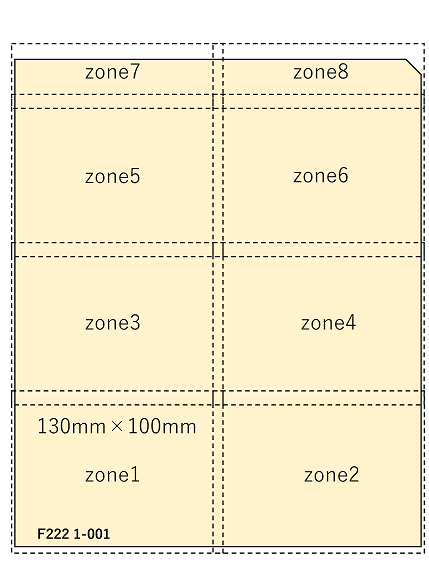}
    \caption{Definition of zones. The films are divided into eight zones for scanning. To determine the orientation of the films, a small triangular section ($\sim$5 mm) is cut at the top-right corner of each film during production.}
    \label{fig:zones}
\end{figure}

The HTS scans a surface of 130 mm $\times$ 100 mm. 
Therefore, a 250 mm $\times$ 300 mm film is divided into eight zones for scanning, as shown in \cref{fig:zones}. 
To ensure complete coverage, adjacent zones overlap by approximately 5 mm. 
It takes approximately 10 minutes to scan each zone, such that both sides of one film can be scanned in approximately 90 minutes, including the time to change the position of the film on the stage.

\section{Reconstruction \label{sec:reconstruction}}



\begin{figure}
    \centering
    \includegraphics[width=0.9\textwidth]{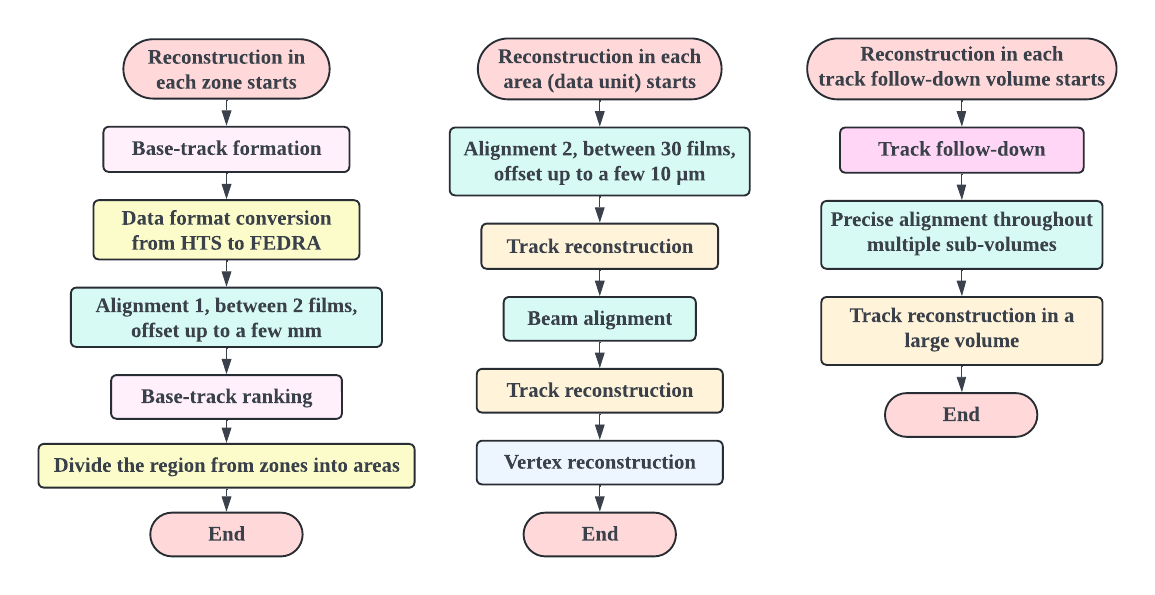}
    \caption{Flowcharts of the reconstruction processes. }
    \label{fig:reconstruction-flow}
\end{figure}

\begin{figure}
    \centering
    \includegraphics[width=0.7\textwidth]{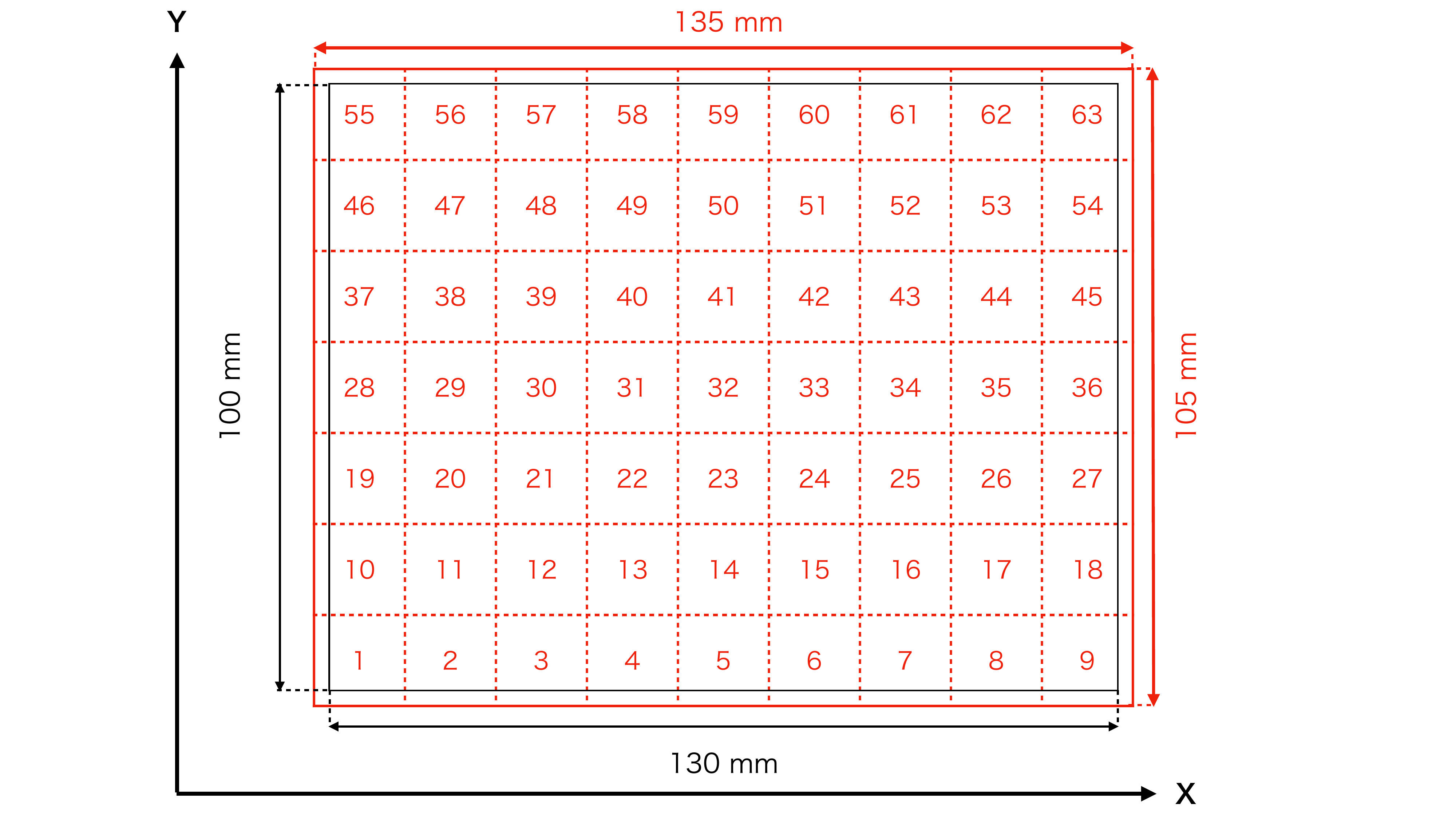}
    \caption{Division of a scanning zone into 63 overlapping areas, each with 1.7 cm $\times$ 1.7 cm size.}
    \label{fig:areas}
\end{figure}

This section details the event reconstruction process using the emulsion detector, as illustrated by the flowcharts in \cref{fig:reconstruction-flow}.
For each scanned zone, track segments passing through the films, which are called base-tracks, are formed from the micro-tracks recorded within the emulsion layers. 
The reconstruction zone is divided into 63 areas, as shown in \cref{fig:areas}.
Each area has a transverse size of $\SI{1.5}{cm} \times \SI{1.5}{cm}$ including $\SI{1}{mm}$ overlap to ensure continuity between adjacent regions. 
Track and vertex reconstruction are performed for data units consisting of 30 plates.
Finally, for tracks near the neutrino vertex candidate, a follow-down method is applied to trace tracks across data units.

\subsection{Base-track formation \label{sec:base-track-formation}}

The HTS system processes a set of 16 images for each emulsion layer to identify micro-tracks.
These micro-tracks, however, include a significant number of fake tracks, which are due to randomly generated silver grains (fog).
In addition, the positions and angles of the micro-tracks have large uncertainties due to the mechanical distortion of the emulsion layer. 

\begin{figure}
    \centering
    \includegraphics[width=0.5\textwidth]{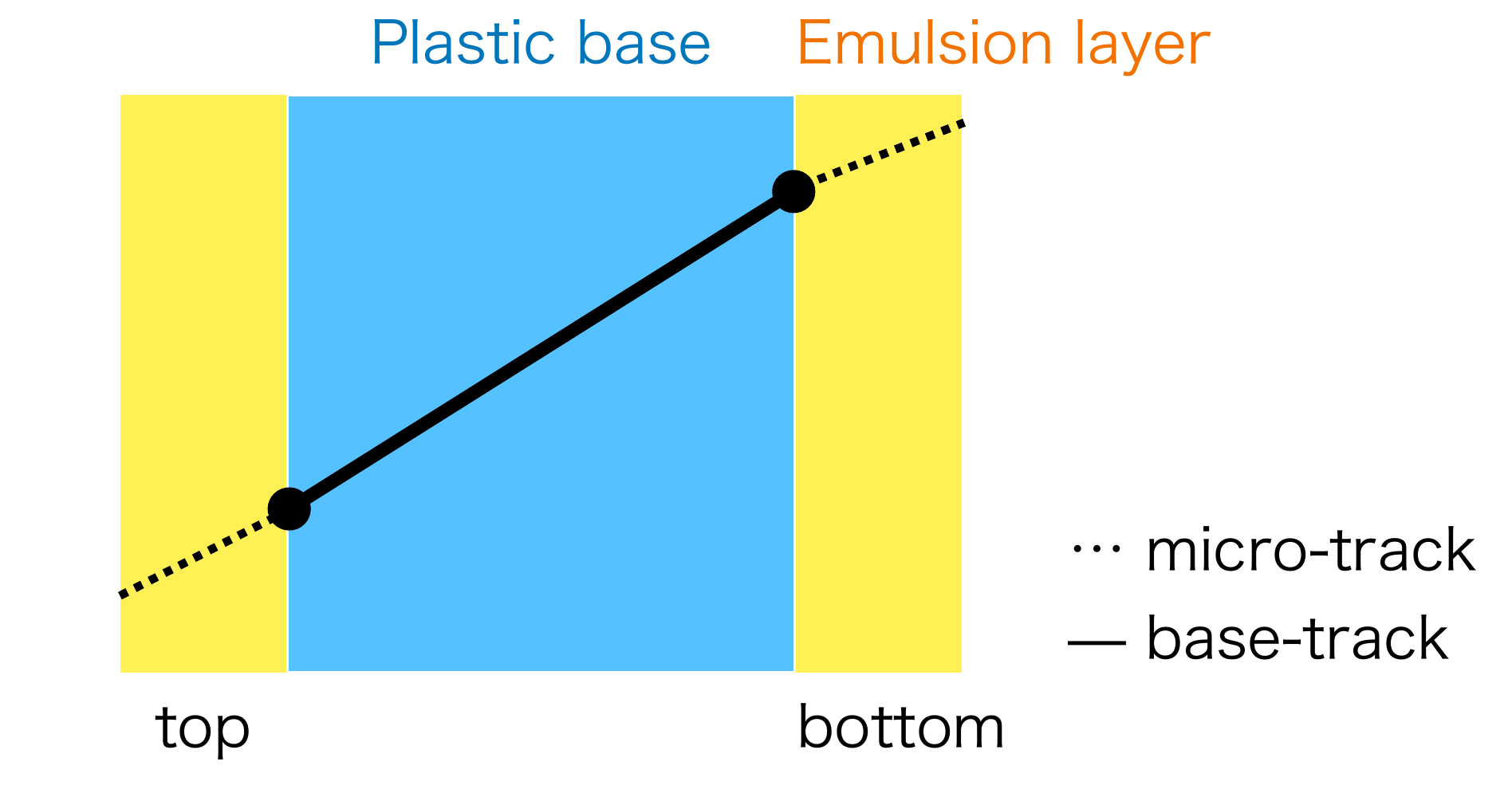}
    \caption{Base-tracks and micro-tracks. The solid lines represent base-tracks, reconstructed from the combinations of the micro-tracks, which are represented by dotted lines.}
    \label{fig:base-track}
\end{figure}

As shown in \cref{fig:base-track}, two micro-tracks, 
each belonging to the top and bottom layers of the film, are paired to form a base-track using their positions and angles. 
The angular measurement of the base-track becomes nearly independent of distortions in individual emulsion layers.  
Consequently, the angular resolution of the base-tracks is approximately ten times better than that of the micro-tracks.

The normalized residuals of the base-tracks $\Delta^{BT}_{MT}$, calculated based on the angular deviations from the micro-tracks, is defined as
\begin{equation}
\textstyle
    \Delta^{BT}_{MT} = \left\{ 
    \left(\frac{\Theta_x^{MT, Top}-\Theta_x^{BT}}{\sigma ( \Theta^{MT}_{x,y})} \right)^2+
    \left(\frac{\Theta_y^{MT, Top}-\Theta_y^{BT}}{\sigma ( \Theta^{MT}_{x,y})} \right)^2+    
    \left(\frac{\Theta_x^{MT, Bot}-\Theta_x^{BT}}{\sigma ( \Theta^{MT}_{x,y})} \right)^2+
    \left(\frac{\Theta_y^{MT, Bot}-\Theta_y^{BT}}{\sigma ( \Theta^{MT}_{x,y})} \right)^2
    \right\}/2,
    \label{eq:delta_btmt}
\end{equation}
where $\tan\theta$ is written as $\Theta$ for simplicity. 
$MT$, $BT$ denote micro-tracks and base-tracks, respectively. 
$Top$ and $Bot$ refer to the top and bottom of the emulsion layer, respectively. 
$\sigma(\Theta^{MT}_{x,y})$ is empirically defined as $\sigma(\Theta^{MT}_{x,y})=0.008 \times \left(1+0.6\times \sqrt{\theta_x^2+\theta_y^2}\right)$, where $\theta_x$ and $\theta_y$ represent the angles of the micro-tracks in the $x$ and $y$ directions, respectively. 
The reconstruction requires that $\Delta^{BT}_{MT} < 15$ for the following reconstruction processes.

\begin{figure}
    \centering
    \includegraphics[width=0.85\textwidth]{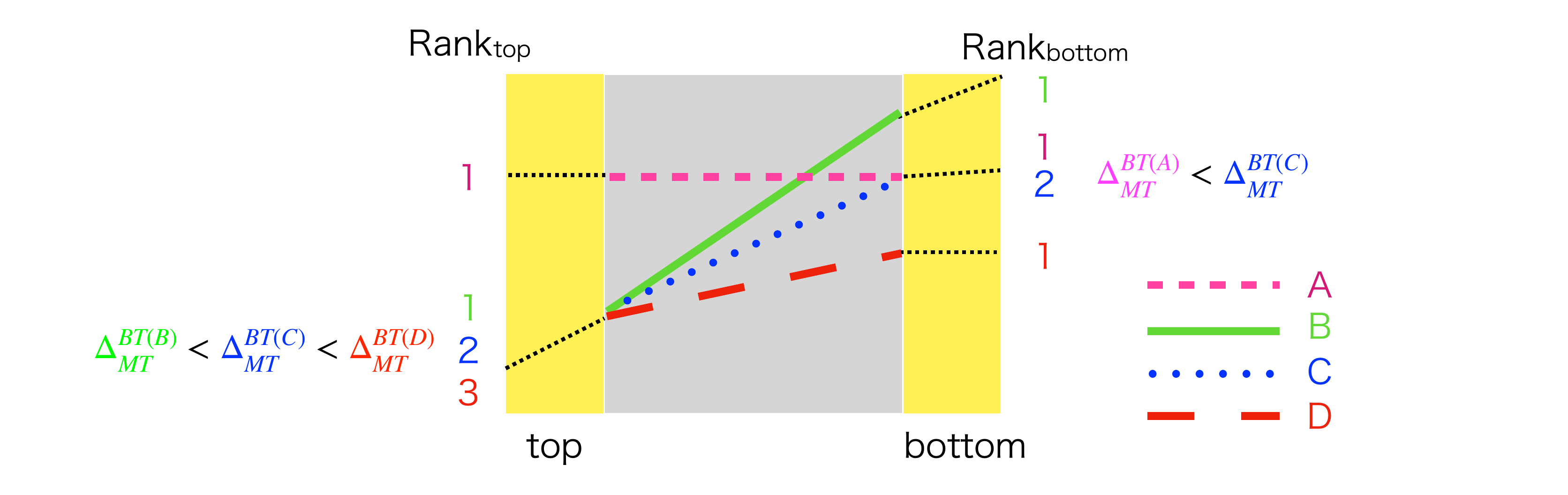}
    \caption{Base-track ranking algorithm. Base-track ranking algorithm is based on $\Delta^{BT}_{MT}$ (\cref{eq:delta_btmt}) when a micro-track connects to multiple base-tracks: the left-bottom micro-track is connected to the three base-tracks (B, C and D). The lower the value of the $\Delta^{BT}_{MT}$, the higher will be the Rank$_\text{top}$ value. The base-tracks with the condition of $\text{Rank}_\text{top} \leq 2$ and $\text{Rank}_\text{bottom} \leq 2$ are selected.}
    \label{fig:bt_ranking}
\end{figure}

Due to the high concentration of tracks recorded in the detector, accidental coincidences of micro-tracks can also produce fake base-tracks. 
This is particularly problematic in the case of electromagnetic showers triggered by high-energy electrons and photons.
This not only degrades the performance of the track reconstruction but also significantly increases the processing time of the track reconstruction. 
To reduce the number of fake base-tracks before the track reconstruction process, base-tracks formed from common micro-tracks are sorted in ascending order based on their $\Delta^{BT}_{MT}$ values, as illustrated in \cref{fig:bt_ranking}.
Up to the best two base-tracks are kept based on this ranking.
This criterion significantly reduces the computing time in the track reconstruction process, while maintaining a high base-track efficiency.

\subsection{Alignment}

Aligning the emulsion films, specifically placing base-tracks from different films within the same coordinate system, is a critical aspect of the FASER$\nu$ event reconstruction. 
Alignment is essential not only for track reconstruction but also for kinematic measurements, such as momentum assessment via multiple Coulomb scattering~\cite{FASER:2019dxq}.

Base-track positions can be displaced for several reasons. 
The mechanical alignment precision of films within each vacuum pack is approximately \SI{100}{\micro\meter} and, between vacuum packs, approximately \SI{500}{\micro\meter}. 
Additionally, the placement of the emulsion films on a microscope during the scanning process introduces another displacement of approximately \SI{500}{\micro\meter}. 
Furthermore, the plastic base of the films is flexible and can be distorted during assembly, development, or placement on the microscope. 
Such distortions result in non-linear displacements that vary with position.

To compensate for these displacements, particle tracks are reconstructed in situ through several steps. 
The alignment parameters between two films are defined using an affine transformation and a gap in the $z$ direction:
\begin{equation}
\left(\begin{array}{c}
     x_{2\,\textnormal{on}\,1}  \\
     y_{2\,\textnormal{on}\,1} 
\end{array}
\right)
=
\left(
\begin{array}{cc}
     a_{2\to 1}&b_{2\to 1}  \\
     c_{2\to 1}&d_{2\to 1} 
\end{array}
\right)
\left(\begin{array}{c}
     x_{2}  \\
     y_{2} 
\end{array}
\right)
+
\left(\begin{array}{c}
     p_{2\to 1} \\
     q_{2\to 1} 
\end{array}
\right),
\end{equation}
\begin{equation}
    z_{2\,\textnormal{on}\,1} = z_2 + \Delta z_{2\to 1}.
\end{equation}
Here, $a_{2\to 1}$, $b_{2\to 1}$, $c_{2\to 1}$, $d_{2\to 1}$, $p_{2\to 1}$, and $q_{2\to 1}$ are constants to be determined. 
$\Delta z_{2\to 1}$ represents the adjustment in the z-coordinate from film 2 to film 1.
This transformation adjusts the coordinates of film 2 to match those of film 1.

There are $\mathcal{O}(10^5)$ muon tracks per cm$^2$, which allow for highly precise alignment. The initial alignment between each pair of consecutive films is performed over a $\SI{13}{\centi\meter} \times \SI{10}{\centi\meter}$ area using recorded base-tracks (alignment 1). 
However, since the emulsion films experience distortion at the $\sim\SI{10}{\micro\meter}$ level across such a large area, a more refined alignment procedure is applied to smaller regions of $\SI{1.5}{\centi\meter} \times \SI{1.5}{\centi\meter}$ (alignment 2), based on the FEDRA framework~\cite{Tyukov:2006ny}. 

\begin{figure}
    \centering
    \includegraphics[width=0.9\linewidth]{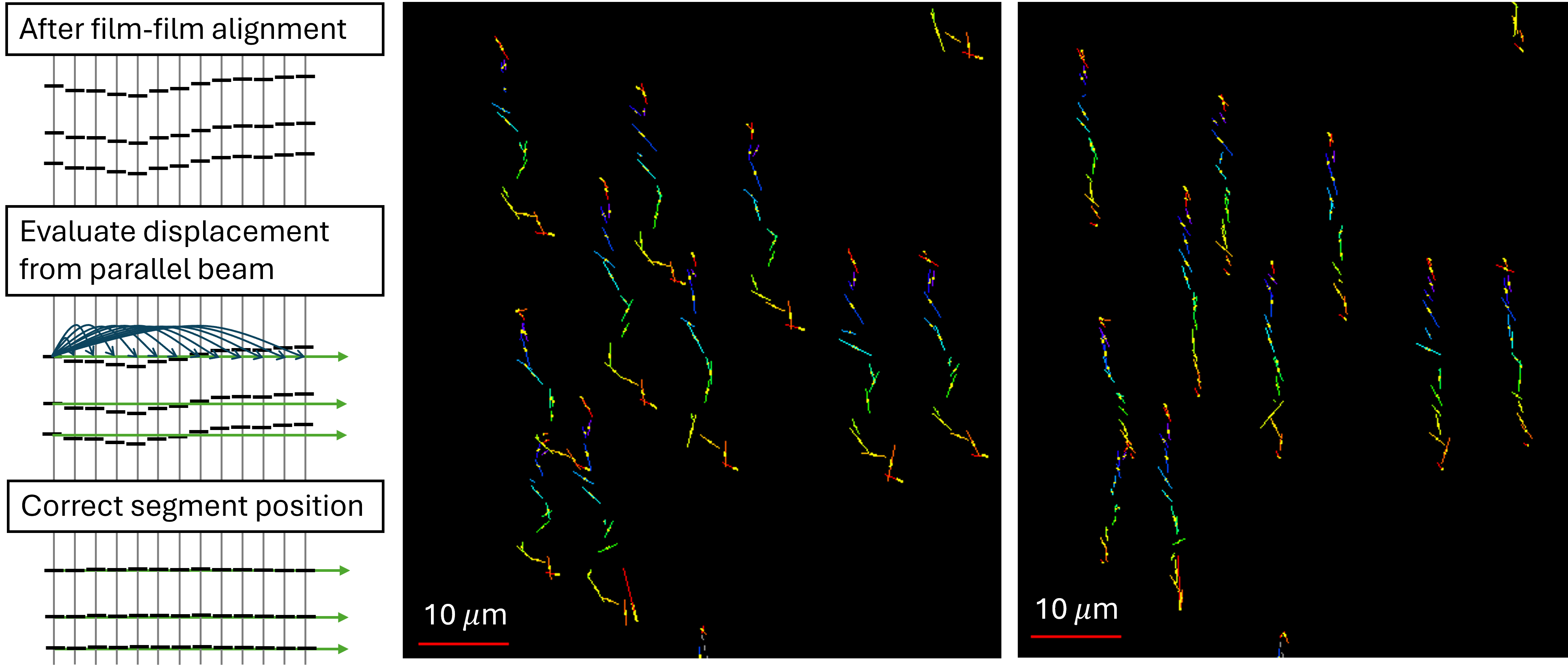}
    \caption{Alignment with a parallel particle beam. Left: a schematic of the beam-alignment procedure. Centre: reconstructed tracks in 15 films without the beam-alignment. Right: the same tracks with the beam-alignment. In the centre and right figures, the beam direction is vertical.}
    \label{fig:finealignment}
\end{figure}

A critical issue of the film-to-film alignment is the accumulation of alignment errors over multiple films. This can cause straight particle tracks to be reconstructed as curved, as shown in the middle panel of \cref{fig:finealignment}. 
To address this, a further alignment is performed on the data-units with 30 plates. 
In this procedure, referred to as beam alignment, all films are aligned relative to the most upstream film, based on the assumption that high-energy muons travel along straight lines. This approach, depicted on the left side of \cref{fig:finealignment}, effectively corrects cumulative alignment errors. As a result, long-range tracks are reconstructed as straight, as shown in the right panel of the figure.
The alignment precision is discussed in~\cref{sec:detector_performance}. 

\subsection{Track reconstruction\label{sec:track_reconstruction}} 

After the first alignment procedure (alignment~1), base-tracks are connected across multiple plates to reconstruct the trajectories of charged particles.  
Note that base-track information is not discarded after track reconstruction, it is used for the following reconstruction processes and the kinematic measurements. 
To align the position displacement of the base-tracks in the localized area (alignment~2), the reconstruction volume is finely segmented before the track reconstruction process.
The segmentation also reduces the computational memory needed for track reconstruction. 


The algorithms for track reconstruction in emulsion detectors used so far are based on a matching test between two base-tracks. 
However, in case of a high track density, such as that expected at the LHC, these algorithms may fail because many tracks arrive at almost the same angle.
Therefore, to cope with the expected very high track density, a new track reconstruction algorithm for emulsion detectors was developed for the NA65/DsTau experiment~\cite{DsTau:2019wjb}. 
This algorithm, also adopted in FASER$\nu$, allows for an improved track reconstruction extending the search for the track combinations across multiple plates.
FASER$\nu$ also performs track reconstruction based on this improved algorithm, making reconstruction in high track density environments possible.

The track reconstruction process requires about 15 minutes per area for 30 plates, for a total data processing of approximately 1 hour per area, including the time for the alignment processes.

\subsection{Vertex reconstruction\label{sec:vertex reconstruction}}

Vertex reconstruction is a crucial step in selecting candidate $\nu_e$ and $\nu_{\mu}$ CC interactions since they form neutral vertices in FASER$\nu$.
Using the reconstructed tracks with $\tan\theta \leq 0.5$ passing through at least three plates, the minimum distances of all two-track combinations are calculated. If this distance is smaller than \SI{3}{\micro m} and the opening angle of the two tracks is larger than 0.03 rad, additional tracks with an impact parameter to the vertex within \SI{5}{\micro m} are searched for. 
Converging patterns with more than four tracks are selected as described in Ref.~\cite{FASER:2024hoe}. Furthermore, converging patterns with more than two tracks are also selected and analysed by requesting an associated high-energy electromagnetic shower with reconstructed energy above 200~GeV to recover efficiency for the events with low track multiplicity. 
The procedure to define an electromagnetic shower is described in Ref.~\cite{FASER:2024hoe}. If multiple vertex candidates are found within \SI{20}{\micro m} in the $x-y$ plane and \SI{300}{\micro m} along the $z$ direction, those with a lower track multiplicity are rejected. 
In the case they have the same track multiplicity, the downstream candidates are rejected.
Tracks are then required to start within three films downstream of the vertices. 
Vertices are categorized as charged or neutral based on the presence or absence, respectively, of charged parent tracks. 
Hits in two more films are required to search for a parent track. 

\subsection{Track follow-down\label{sec:track_followdown}}
The tracks reconstructed in individual sub-volumes, whose length do not exceed 30 plates, are used to search for vertices.
For muon identification and further analysis, a follow-down process for tracks (track follow-down, TFD) lying in the same area within $\pm$0.5 cm from the reconstructed vertex is performed. 
Thus, the maximum transverse dimension of the TFD volume is \SI{1}{cm^2}. Tracks reconstructed in the data units are associated with one another, and those left by the same particles are connected from upstream to downstream. 
Since subsequent sub-volumes, which represent a unit of reconstruction in the beam direction, overlap, a track reconstructed in the upstream sub-volume can share several common segments with the corresponding track reconstructed in the next downstream sub-volume. 
Common segments are unambiguously distinguished by their identifiers, which are unique for each plate. 
Two tracks are then connected if they have at least six common segments. 
The base-track reconstruction efficiency of the final tracking results was verified to be independent of the selected value, ranging from six to ten common segments. 

After two tracks from consecutive sub-volumes are associated to each other, segments of the track from the downstream sub-volume starting from the first non-common plate, are added to the corresponding track from the 
upstream sub-volume.

All the other tracks not connected from the downstream sub-volume are also added to the upstream sub-volume if they have more than one segment after the first non-common plate of the downstream sub-volume. Segments of such tracks are removed if they belong to one of the 20 common plates of the two overlapping sub-volumes. 
Part of these tracks can be connected with tracks from the downstream sub-volumes. 
Other, short non-connected tracks are used in further analysis of neutrino event topologies, such as reconstruction of secondary hadron interactions or decay vertices.

After all analysed sub-volumes are connected, a more precise alignment (through multiple sub-volumes) is performed using tracks with a number of segments larger than 11. 
Segments from the first ten plates are used to estimate the track direction. 
The approximated $x$-$z$ and $y$-$z$ track slopes are used instead of individual track segment slopes for the alignment of consecutive plates. 
The obtained affine parameters, as well as the differences in the $z$ positions, are used for the correction of all the track segments saved for further analysis. 
Finally, the tracking procedure is performed once again through all the connected sub-volumes by means of the aligned segments.

\begin{figure}
    \centering
    \includegraphics[width=0.8\linewidth]{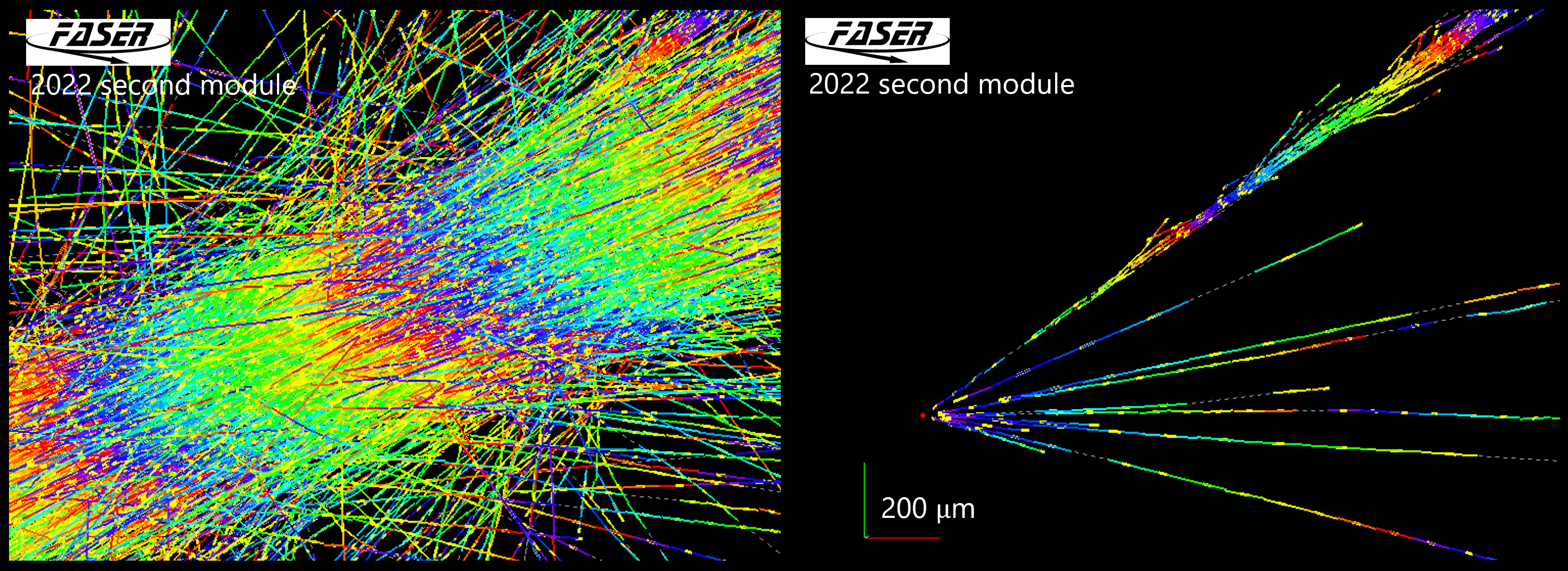}
    \caption{Example of event displays in the same 3D view showing all the reconstructed tracks in a region (left) and the reconstructed tracks associated with the neutrino vertex (right). 
    In the right panel, the right-handed coordinate axes are shown in the bottom left, with blue, green, and red axes indicating the $x$ (horizontal), $y$ (vertical), and $z$ (beam) directions, respectively. Yellow line segments show the trajectories of charged particles in the emulsion films. The other coloured lines are interpolations, with the colours indicating the longitudinal depth in the detector. 
    The data corresponds to one of the electron neutrino events reported in Ref.~\cite{FASER:2024hoe}. The displays show the same data region, demonstrating the ability to reconstruct a significant number of accumulated tracks and finally observe neutrino interactions. The longest track shown is going through 82 plates.}
    \label{fig:event-display}
\end{figure}

\cref{fig:event-display} shows a typical event display with tracks reconstructed in a track follow-down volume and tracks associated with the reconstructed neutrino vertex. 
It demonstrates the ability to reconstruct a significant number of accumulated tracks and observe neutrino interactions.

\section{Evaluation of Detector Performance  \label{sec:detector_performance}}
The performance of the FASER$\nu$ emulsion detector has been evaluated using data collected between July 26 and September 13, 2022, corresponding to 9.5 fb$^{-1}$ of $pp$ collisions at a centre-of-mass energy of \SI{13.6}{TeV}. 
This dataset provides a comprehensive basis for assessing various detector characteristics including track reconstruction, single-film efficiency, and position and angular resolution. 
The results demonstrate that the detector performs as designed and is capable of achieving the required precision for neutrino interaction measurements~\cite{FASER:2020gpr}. 
The evaluation is supported by detailed simulations where neutrino interactions are modeled using \texttt{GENIE}~\cite{Andreopoulos:2009rq,Andreopoulos:2015wxa} while all the other interactions are simulated using \texttt{GEANT4}~\cite{GEANT4:2002zbu}. 
These simulations incorporate realistic detector geometry and apply corrections to account for alignment and efficiency. 
Both simulation and experimental data are reconstructed using the same method for a consistent comparison.
The following subsections detail specific aspects of the detector performance evaluation.

\subsection{Track density}

\begin{figure}[]
\centering
\includegraphics[width=0.45\linewidth]{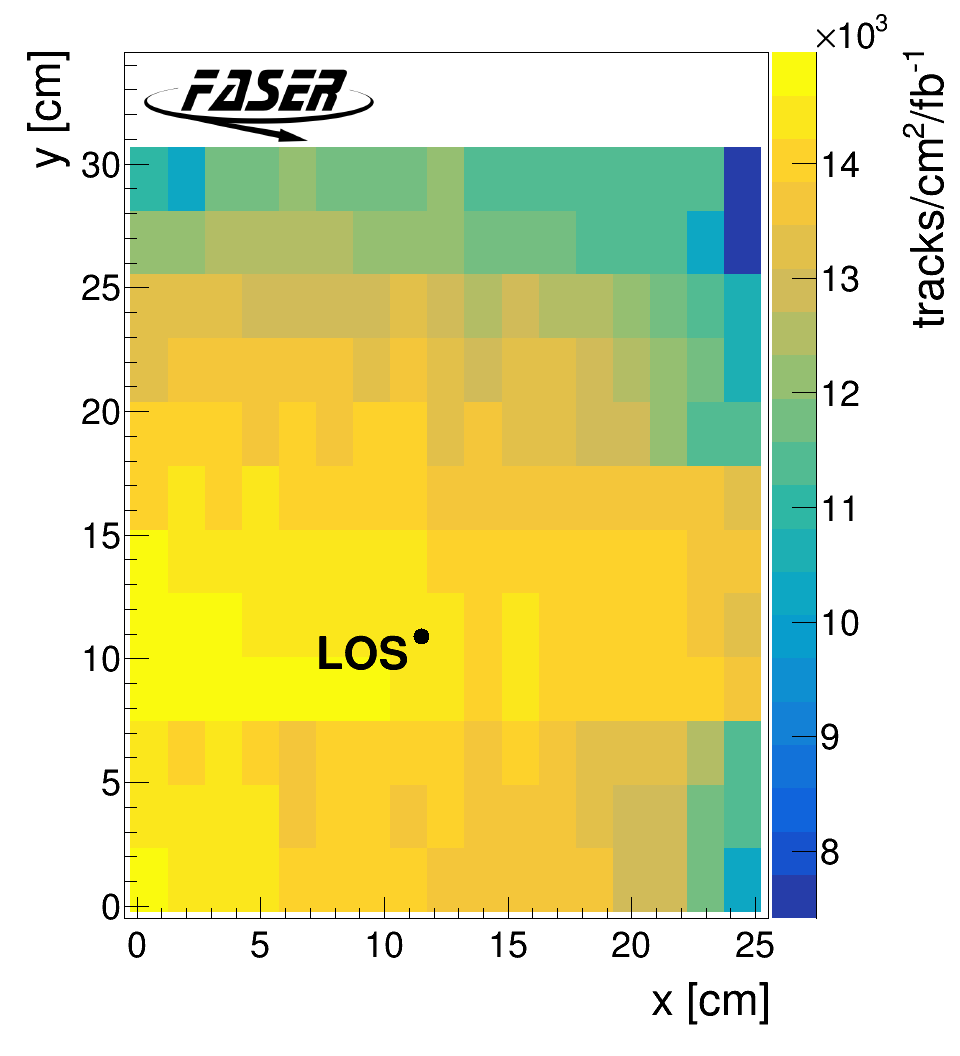}
\caption{Track density map in the sub-volume from the detector plate 41 to 70. 
Tracks starting within the first five films and having more than five track segments and a track angle within 10 mrad from the angular peak are counted. 
Track angles are calculated by connecting the position of the first and the last segment of the tracks. 
The number of tracks is divided by 9.5 fb$^{-1}$.
The  line of sight (LOS) point represents the beam collision axis in the emulsion detector. }
\label{fig:track_density}
\end{figure}

The FASER$\nu$ emulsion detector, due to its intrinsic lack of time information, records all charged particles passing through it. 
Muons from the LHC collisions provide the main contribution through the detector. 
\cref{fig:track_density} shows the track density distribution reconstructed in the data-units from the detector plate 41 to 70. 
No efficiency correction is applied to the track density calculation.
A high density of $\mathcal{O}(10^5)$ tracks per $\si{cm^2}$ has been measured across the bulk of the films, excluding the edges.
For a track density of this order of magnitude, the average distance between adjacent tracks is approximately $\SI{32}{\mu m}$.
The observed track density has a weak dependency on the position.

\subsection{Single-film efficiency}
\begin{figure}[]
\centering
\includegraphics[width=0.4\linewidth]{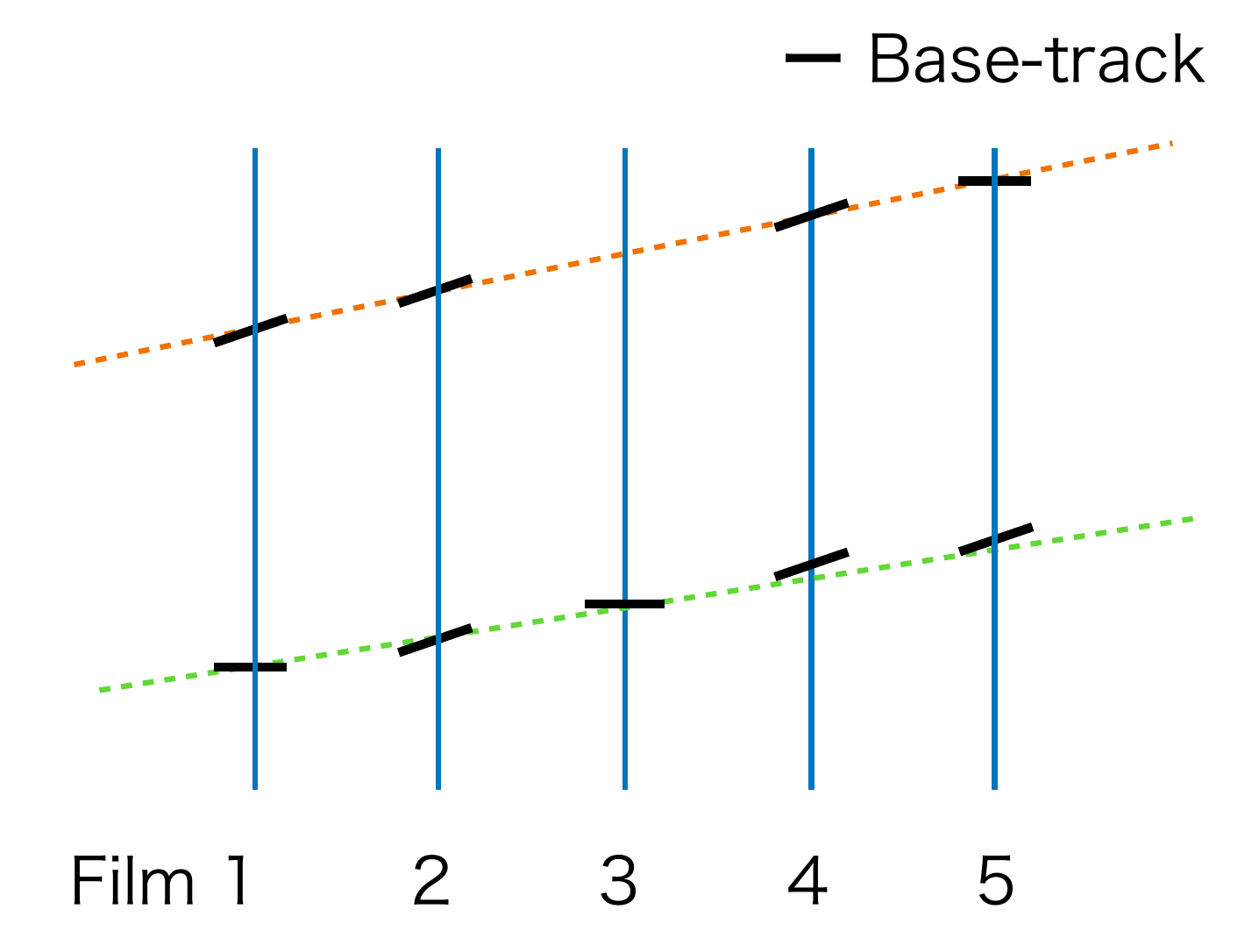}
\caption{Patterns used for the calculation of a single-film efficiency. It is defined as the ratio of the number of tracks with hits recorded in the measured film (film 3) to the total number of tracks passing through it. Specifically, the denominator is the number of tracks with hits recorded in the two upstream and two downstream plates of the measured film (film 1, 2, 4, and 5). The numerator is the number of those tracks that also have hits recorded in the measured film.}
\label{fig:EfficiencyCalculation}
\end{figure}

\begin{figure}[]
\centering
\includegraphics[width=0.45\linewidth]{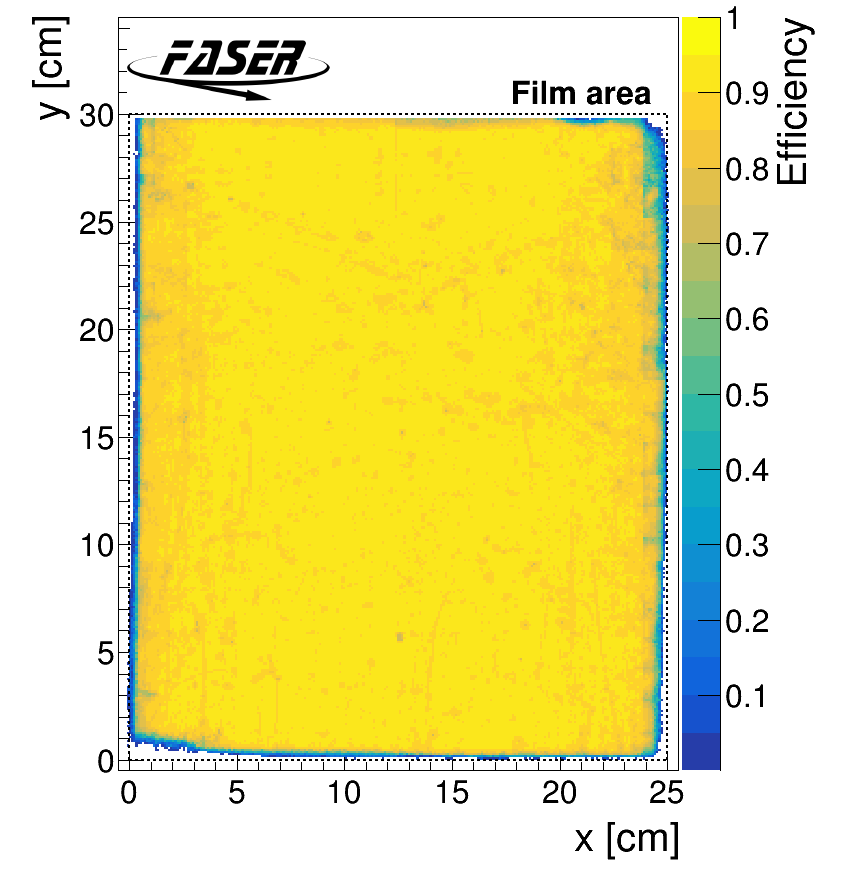}
\caption{Distributions of the single-film efficiency for the sub-volumes consisting of plates from 41 to 70 of the detector. }
\label{fig:efficiency}
\end{figure}

The single-film efficiency is used to check the quality of the films, selecting the regions to be used in the physics analysis, and determining the efficiency correction parameters for the Monte Carlo simulations.
\cref{fig:EfficiencyCalculation} shows how the efficiency is assessed. 
It is calculated using tracks reconstructed with five plates centred on the plate of interest. 
The efficiency is defined as the fraction of such tracks which have hits in this plate. 
As shown in \cref{fig:efficiency}, an efficiency of more than 85\% is observed across the bulk of the films, with the exception of the edges, where regions with efficiency below 80\% are observed. 
Therefore these regions are excluded in the physics analysis. 
Typical track requirements on the reconstructed tracks used for neutrino reconstruction require three hits out of seven films, which leads to a tracking efficiency of approximately 99.5\% for a single-film efficiency of 80\%.

\subsection{Position and angular resolution}
\begin{figure}[]
\centering
\includegraphics[width=0.3\linewidth]{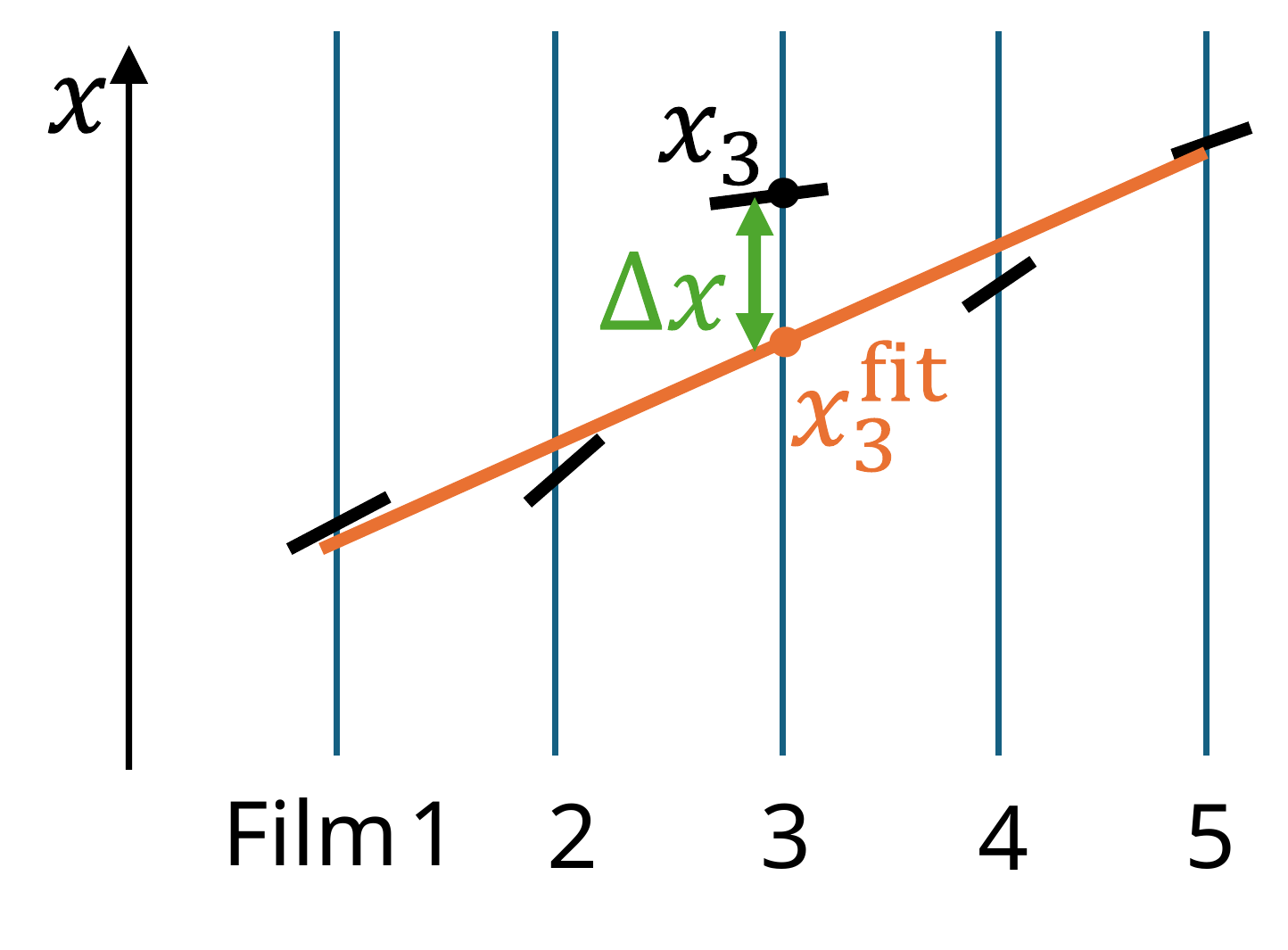}
\includegraphics[width=0.3\linewidth]{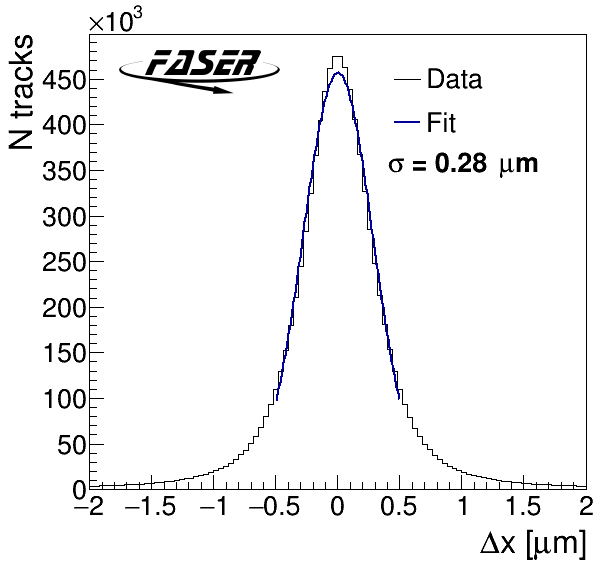}
\includegraphics[width=0.3\linewidth]{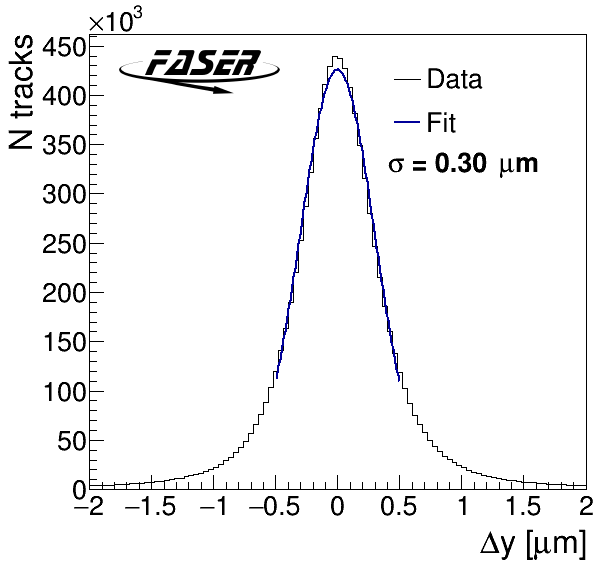}
\caption{Evaluation of the alignment accuracy. Left: a schematic of the method. Centre and right: distributions of the position deviation of the base-tracks from the fitted line in a volume after the track follow-down processes. Gaussian fittings (blue lines) give an alignment resolution of \SI{0.3}{\micro m}.
}
\label{fig:position_resolution}
\end{figure}

\begin{figure}
    \centering
    \includegraphics[width=0.6\linewidth]{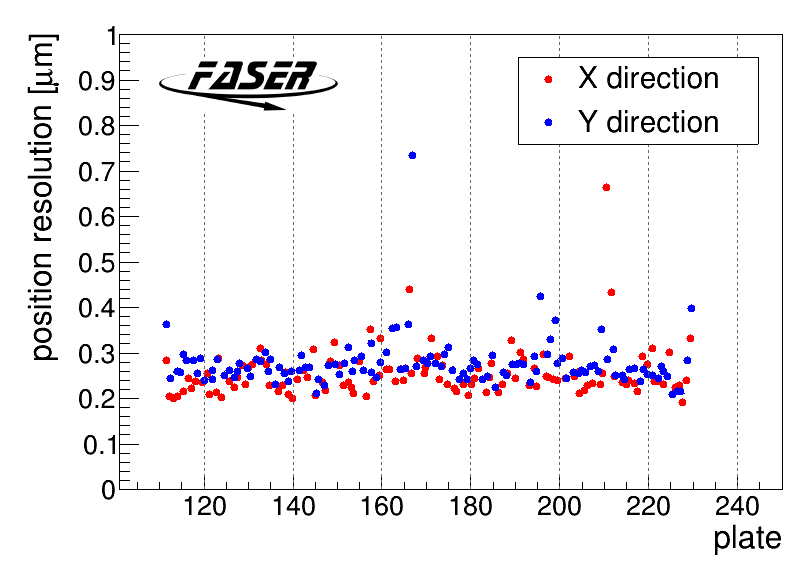}
    \caption{Position resolution as a function of plate number for one of the track follow-down volumes (plates 111 to 220).}
    \label{fig:position_resolution_plate}
\end{figure}

The position accuracy of the emulsion detector is essential for both the kinematic analysis of neutrino interactions and the measurement of individual charged particle momenta. 
A position resolution of \SI{0.4}{\micro m} is required for measuring a momentum of \SI{1}{TeV} based on the multiple Coulomb scattering method.
The position resolution, or alignment precision, is evaluated using sets of five emulsion films. 
As shown schematically in \cref{fig:position_resolution} on the left, a straight line is fitted using the base-tracks on films 1, 2, 4, and 5, and the deviation of the base-track on film 3 from the fitted line is evaluated. 
As shown in \cref{fig:position_resolution} on the right, the typical position resolution is found to be \SI{0.3}{\micro m}. 
\cref{fig:position_resolution_plate} shows the position resolution as a function of the plate number.
Relatively poor resolution is observed in some plates, which is due to local distortion of the emulsion films. 

Track angle measurement is a crucial component for both topological and kinematical analyses. The angles of leptons and hadrons, as well as their difference in the transverse plane with respect to the neutrino direction, $\Delta\phi$, are important variables for selecting neutrino events from neutral hadron backgrounds and determining the neutrino energy~\cite{FASER:2024hoe}.

\begin{figure}
    \centering
    \includegraphics[width=0.45\linewidth]{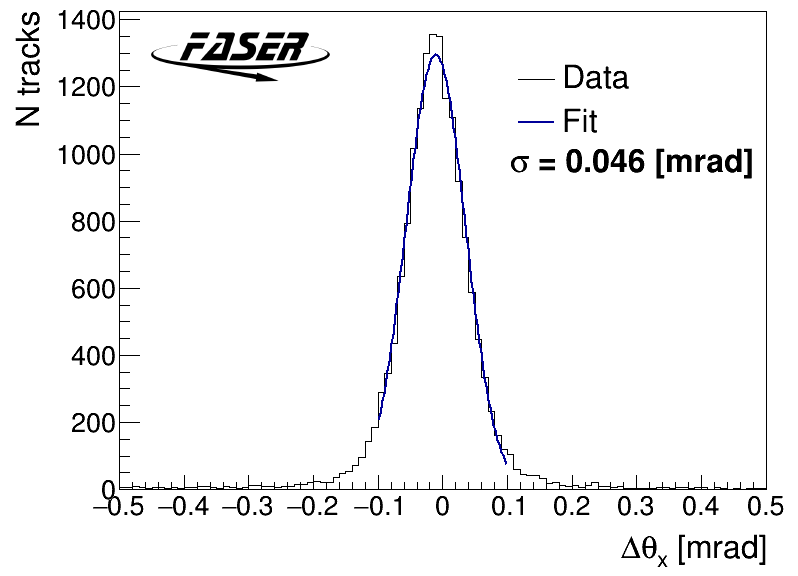}
    \includegraphics[width=0.45\linewidth]{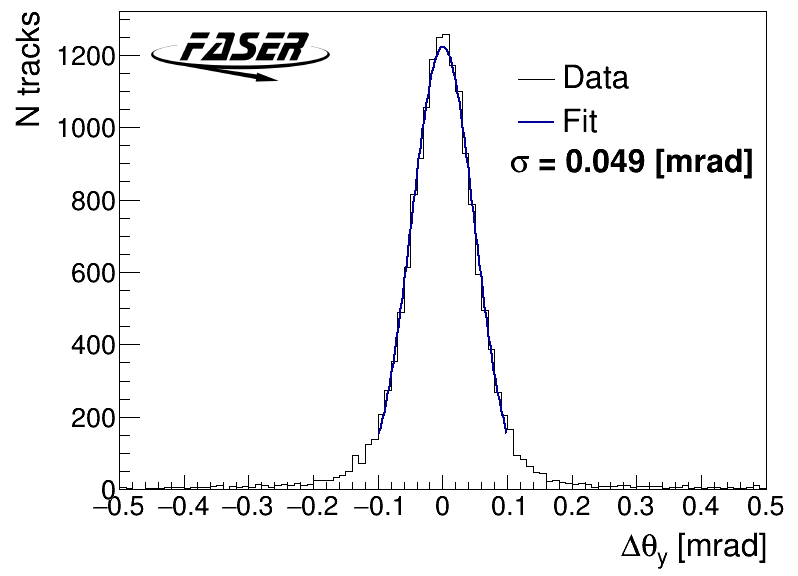}

    \caption{Angular resolution measured in data. See the text.}
    \label{fig:dtx}
\end{figure}

The angular resolution of the reconstructed tracks is evaluated by using background muons. The angular distribution of the muons has a spread of approximately \SI{400}{\micro rad}, corresponding to momenta above 1~TeV, considering the scattering in the 100~m of rock located in front of FASER$\nu$. Track angles are calculated by analyzing the first 10 track segments and the subsequent 10 segments within a track-follow-down volume. The angular difference between these two calculations is shown in \cref{fig:dtx}. 
The Gaussian sigma of the resulting distribution was determined to be \SI{46}{\micro rad} in the $x$-direction and \SI{49}{\micro rad} in the $y$-direction.


\subsection{Impact parameter}

The impact parameter, defined as the minimum distance between the extrapolated track and the reconstructed vertex position, is a crucial quantity for evaluating the vertex reconstruction accuracy. 
A smaller impact parameter indicates better alignment between the tracks and the vertex, which is essential for identifying neutrino interaction events with high precision.

\begin{figure}[h]
\centering
\includegraphics[width=0.3\linewidth]{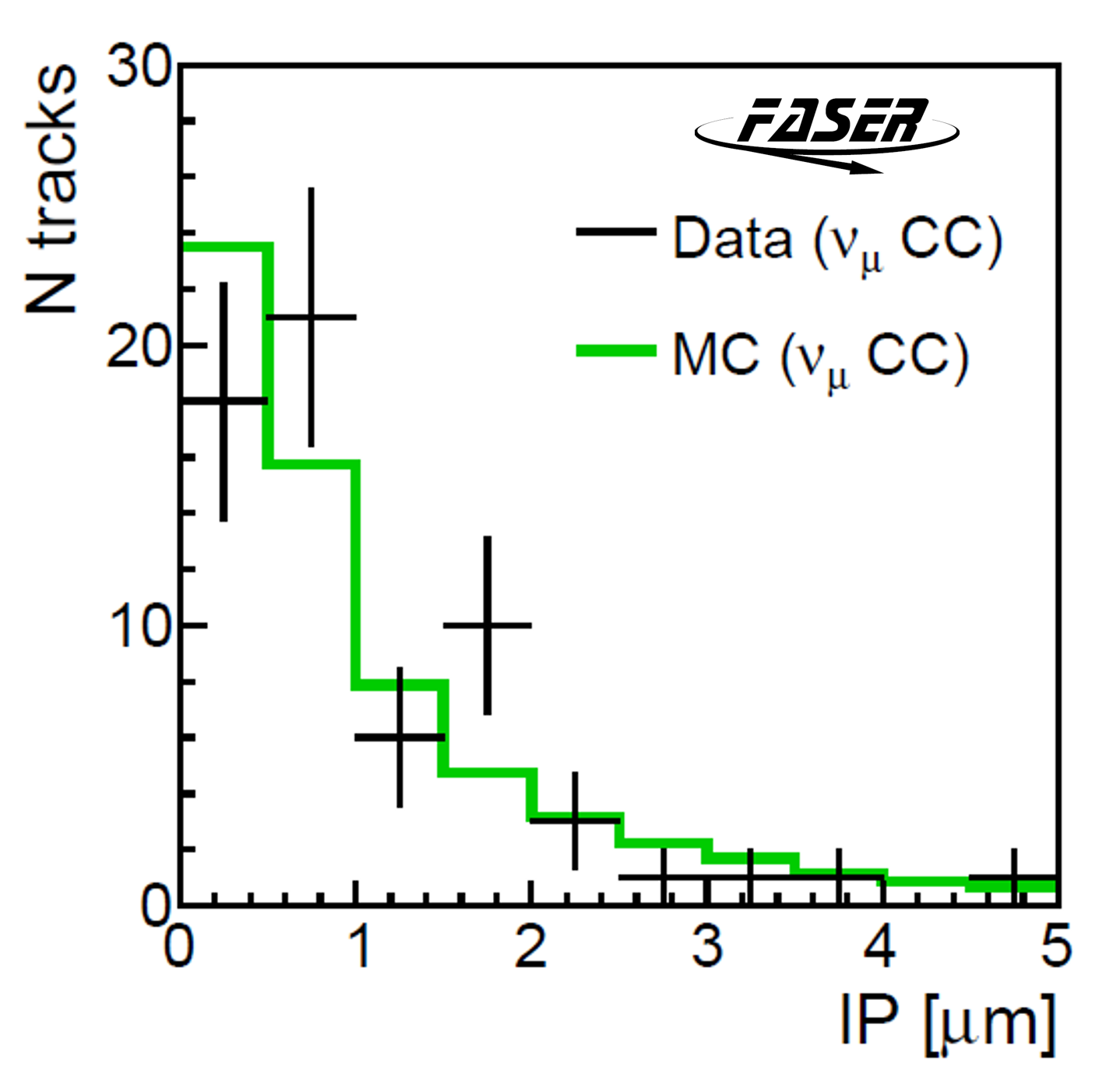}
\caption{The distribution of the impact parameter of the tracks forming the vertices with respect to the reconstructed vertex positions for the observed $\nu_{\mu}$ signal events reported in~\cite{FASER:2024hoe}. The observed $\nu_{\mu}$ signal events from a 128.8 kg subset of the FASER$\nu$ volume, which collected 9.5 fb$^{-1}$ of data, are shown in black. The distribution from Monte Carlo simulation is represented in green. The simulated distribution is normalized to the number of observed tracks.}
\label{fig:ip_numuCC}
\end{figure}

The vertex reconstruction algorithm discussed in \cref{sec:vertex reconstruction} reconstructs about 52\% (45\%) of $\nu_e$ ($\nu_{\mu}$) CC interactions. 
Out of the reconstructed neutral vertices in data, about 5\% pass the neutrino event selection described in~\cite{FASER:2024hoe}, while the remaining vertices are mainly due to neutral hadron interactions. 
\cref{fig:ip_numuCC} shows the distribution of the impact parameter of the tracks forming the vertices to the reconstructed vertex positions for the $\nu_{\mu}$ CC signal events reported in Ref.~\cite{FASER:2024hoe}.
The agreement between the observed and simulated impact parameter distributions demonstrates the good performance of the vertex reconstruction algorithm. 
This comparison also serves as a validation of the simulation framework used for $\nu_{\mu}$ CC signal modeling.

\section{Conclusion \label{sec:summary}}
This paper presents the event reconstruction techniques and performance evaluation of the FASERν emulsion detector at the LHC, which achieves sub-micron precision in the reconstruction of charged particle tracks produced in the LHC collisions. 
The detector, constructed from 730 layers of tungsten and emulsion films, enables precise measurements of neutrino interactions. The challenges posed by high-density track environments, reaching $\mathcal{O}(10^5)$ tracks per $\si{cm^2}$, were addressed with advanced alignment techniques, dedicated track and vertex reconstruction algorithms, and a follow-down method for linking tracks across sub-volumes.
Algorithms for track reconstruction at even higher densities are under development and are expected to be implemented in the future.

The data processing time for the described detector structure can be very roughly estimated for each zone as approximately one day/one plate/(one CPU $+$ one GPU). 
In case of four processing servers with six processes running in parallel on each of them, it takes about seven months for full data processing of one FASERν module.

The performance of the detector is evaluated in detail. 
More than 85\% single-film efficiency is achieved across most regions of the emulsion films. 
The detector has an alignment precision of $\SI{0.3}{\mu m}$, which is sufficient for accurate momentum and kinematic studies. 
High track densities are successfully handled across the detector area, demonstrating the robustness of the system in a complex experimental environment. 
These results highlight the detector’s capability to identify neutrino interactions with exceptional spatial resolution and establish a solid base for future studies using the FASER$\nu$ emulsion detector.

\section*{Acknowledgments}
\label{sec:Acknowledgments}
\input{acknowledgments}

\appendix


\bibliographystyle{utphys}
\bibliography{references}

\end{document}

%% file: authorlist.tex
\author[1]{Roshan Mammen Abraham\,\orcidlink{0000-0003-4678-3808},}
\author[2]{Xiaocong Ai\,\orcidlink{0000-0003-3856-2415},}
\author[3]{Saul Alonso Monsalve\,\orcidlink{0000-0002-9678-7121},}
\author[4]{John Anders\,\orcidlink{0000-0002-1846-0262},}
\author[5]{Claire Antel\,\orcidlink{0000-0001-9683-0890},}
\author[6,7]{Akitaka Ariga\,\orcidlink{0000-0002-6832-2466},}
\author[8]{Tomoko Ariga\,\orcidlink{0000-0001-9880-3562},}
\author[6]{Jeremy Atkinson\,\orcidlink{0009-0003-3287-2196},}
\author[9]{Florian U. Bernlochner\,\orcidlink{0000-0001-8153-2719},}
\author[9]{Tobias Boeckh\,\orcidlink{0009-0000-7721-2114},}
% ...

\affiliation[1]{Department of Physics and Astronomy, University of California, Irvine, CA 92697-4575, USA}
\affiliation[2]{School of Physics, Zhengzhou University, Zhengzhou 450001, China}
\affiliation[3]{ETH Zurich, 8092 Zurich, Switzerland}
\affiliation[4]{CERN, CH-1211 Geneva 23, Switzerland}
\affiliation[5]{D\'epartement de Physique Nucl\'eaire et Corpusculaire, University of Geneva, CH-1211 Geneva 4, Switzerland}
\affiliation[6]{Albert Einstein Center for Fundamental Physics, Laboratory for High Energy Physics, University of Bern, Sidlerstrasse 5, CH-3012 Bern, Switzerland}
\affiliation[7]{Department of Physics, Chiba University, 1-33 Yayoi-cho Inage-ku, 263-8522 Chiba, Japan}
\affiliation[8]{Kyushu University, 744 Motooka, Nishi-ku, 819-0395 Fukuoka, Japan}
\affiliation[9]{Universit\"at Bonn, Regina-Pacis-Weg 3, D-53113 Bonn, Germany}
% ...

%% file: acknowledgments.tex
We thank CERN for the excellent performance of the LHC and the technical and administrative staff members at all FASER institutions for their contributions to the success of the FASER experiment. We thank the ATLAS Collaboration for providing us with accurate luminosity estimates for the Run 3 LHC $pp$ collision data. We also thank the CERN STI group for providing detailed \texttt{FLUKA} simulations of the muon fluence along the LOS, the CERN EN-HE group for their help with the installation/removal of the FASER$\nu$ detector, and the CERN EP department for the refurbishment of the CERN dark room, used for the preparation and development of the FASER$\nu$ emulsion. We thank Saya Yamamoto for supporting the preparation of emulsion films. 

This work was supported in part by Heising-Simons Foundation Grant Nos.~2018-1135, 2019-1179, and 2020-1840, Simons Foundation Grant No.~623683, U.S. National Science Foundation Grant Nos.~PHY-2111427, PHY-2110929, and PHY-2110648, JSPS KAKENHI Grant Nos.~19H01909, 22H01233, 20K23373, 23H00103, 20H01919, and 21H00082, the joint research program of the Institute of Materials and Systems for Sustainability, ERC Consolidator Grant No.~101002690, BMBF Grant No.~05H20PDRC1, DFG EXC 2121 Quantum Universe Grant No.~390833306, Royal Society Grant No.~URF$\backslash$R1$\backslash$201519, UK Science and Technology Funding Councils Grant No.~ST/ T505870/1, the National Natural Science Foundation of China, Tsinghua University Initiative Scientific Research Program, and the Swiss National Science Foundation.